\documentclass[11pt,a4paper]{article}
\usepackage{amssymb}
\usepackage{amsmath}
\usepackage{amsfonts}
\usepackage[mathcal]{eucal}
\usepackage[english]{babel}
\usepackage{amscd}
\usepackage{xypic}
\newcommand{\ks}{Kustaanheimo-Stiefel }
\newcommand{\rt}{\mathbb{R}^{3}_{\scriptscriptstyle{0}}}
\newcommand{\rqq}{\mathbb{R}^{4}_{\scriptscriptstyle{0}}}

\newcommand{\emme}{\CMcal{M}}
\newcommand{\enne}{\CMcal{N}}
\newcommand{\A}{`}
\newcommand{\FM}{\CMcal{F}\left(\CMcal{M}\right)}
\newcommand{\FN}{\CMcal{F}\left(\CMcal{N}\right)}
\newcommand{\Frt}{\CMcal{F}\left(\rt\right)}
\newcommand{\Frqq}{\CMcal{F}\left(\rqq\right)}
\newcommand{\de}{\mathrm{d}}
\newcommand{\Di}{\mathrm{D}}
\title{Reduction and Unfolding for Quantum Systems: the Hydrogen
Atom}%
\author{Antonella D'Avanzo, Giuseppe Marmo, Alessandro
Valentino\\[3mm]
\textit{Dipartimento di Scienze Fisiche, Universit\'a Federico II}\\
\textit{and}\\
\textit{Istituto Nazionale di Fisica Nucleare - Sezione di
Napoli}\\[2mm]
davanzo@na.infn.it, marmo@na.infn.it, valentino@na.infn.it}%
\date{}
\begin{document}%
\maketitle

\begin{abstract}
\noindent In this paper we propose a ``quantum reduction
procedure'' based on the reduction of algebras of differential
operators on a manifold. We use these techniques to show, in a
systematic way, how to relate the hydrogen atom to a family of
quantum harmonic oscillators, by the means of the
Kustaahneimo-Stiefel fibration.
\end{abstract}
\section{Introduction}

Reduction procedures have been extensively studied in the
classical setting, dealing mainly with Hamiltonian dynamics
(%\cite{whitt},
\cite{MW74}, \cite{AM}, \cite{MSSV}) but also Lagrangian dynamics
or more general ones (\cite{red},\cite{red1}, \cite{genred}). It
has been noted that different classes of completely integrable
systems arise as reduction of free (or simpler) ones in higher
dimensions (\cite{pere}): this has been the motivation for the
development of unfolding techniques, i.e. the converse of
reduction, a still open field. In this context some of us (A.D.,
G.M.) have studied in a systematic way how to relate the Kepler
problem to a family of harmonic oscillators (\cite{ruclass}); the
aim of the present paper is to exploit this same relationship in
the quantum setting, focusing on a possible method of quantum
reduction that we propose here and on a possible way to tackle
the unfolding procedure.\\
The problem we consider as a case of study has been widely studied
and the above relationship has been established using many
different approaches (\cite{pauli},\cite{boiteaux},
\cite{chen},\cite{gerry},\cite{gracia},\cite{path},\cite{cordani2});
so, the results are well known (\cite{Wyb}, \cite{englefield}).
However, all these approaches lack of a systematic procedure
underlying the computational results: moved by this, in this paper
we try to insert the problem related to the hydrogen atom in a
more general framework, i.e. that of ``quantum reduction''.
Although the problem of reduction has been given a lot of
consideration in the classical setting, the same has not happened
in a systematic way in the quantum setting. An attempt toward this
direction has been done by Perelomov and Olshanetsky in
\cite{pere2}. This article provides examples and motivations to
the idea that it could be useful to develope an unfolding
procedure also in the quantum setting. Some other proposals are
given in \cite{guillstern}, \cite{KKS}, and also \cite{lieconstr}.
Within this context, the present paper try to clarify some
methodological aspects of these procedures, that are neverthless
not yet well systematized, and to make a further step
towards the development of a \A\A{}quantum'' reduction.\\
Our proposal has its starting point in the analysis of the
classical reduction procedure, which, in an algebraic language
(\cite{genred}), can be seen as a homomorphism between selected
Lie algebras of vector fields on manifolds, associated with a
suitable map between the configuration manifolds. In the same
spirit, we propose a \A\A quantum reduction'' procedure based on a
homomorphism between Lie algebras of differential operators (of
arbitrary degree) on manifolds. Our choice to deal with
differential operators built on the \A\A configuration space'' has
been made to emphasize the geometrical aspects of a quantum
mechanical system, and to work in strict analogy with the
classical case. Nevertheless, we treat the problem as a purely
quantum one, without resorting to any kind of quantization.\\
For the sake of clarity, we decompose the problem in two steps.
The first one is the development of a reduction procedure for
differential operators on a manifold $\CMcal{M}$, i.e. acting on
the  functions on $\CMcal{M}$, without dealing with the Hilbert
space structure, that will be considered in the second step. The
first step could be thought of as the reduction of a partial
differential equation on a manifold to one on a manifold of lower
dimensions. The definitions and results we propose are therefore
not necessarily linked to quantum mechanics, since they concern
only differential equations, so they can be used in several physical contests. \\
However, since we are interested in quantum mechanics, the Hilbert
space structure plays a relevant role and has to be taken into
account. However, there is  not an algoritmic way to build the
relationship between Hilbert spaces: it strongly depends on the
features of the map between the configuration manifolds,
and this has to be analyzed case by case. \\
All this has been worked explicitely in the case of the hydrogen
atom, where we have been able to restrict the arbitrariness in the
choice of the unfolding operator, by means of an ``educated
guess'' motivated by symmetry considerations. Our procedure
emphasizes the relationship of the hydrogen atom with a
one-parameter
\textit{family} of harmonic oscillators.\\
Motivated by what happens in our case of study, we conclude with
an attempt toward the analysis of the possible role of
reparametrization in the quantum setting. As far as we know, this
topic has never been fully exploited: in our opinion, this would
shed more light on the relationship between classical and quantum
settings, both physically and mathematically.\\
The paper is organized as follows.\\
In section 2, after recalling a way to deal with differential
operators (of any degree) on a manifold, we introduce and motivate
the definition of \A\A projectable differential operator''. In the
ending subsection we give some additional details about the
projectability of differential operators on
$\mathbb{R}^{4}-\{0\}\equiv\rqq$ on differential
operators on $\mathbb{R}^{3}-\{0\}\equiv\rt$ with respect to the so-called \ks map.\\
In section 3 we apply these techniques to the hydrogen atom,
introducing a possible \A\A{}quantum reduction procedure''. Our
approach naturally provides the relation with a one parameter
family of harmonic oscillators, as in the classical case.
Moreover, we are able to recover usual results about the
eigenvalue problem and the algebra of simmetry (section 3.1 and
3.2 respectively).\\
In the last section, we comments upon the role of the
\A\A{}reparametrization'', which was an important feature of the
classical case, also in the quantum setting.
 \numberwithin{equation}{section}
\section{Differential operators}
The aim of this section is to give a short review to recall how to
deal with differential operators on a manifold. In the following
we will restrict our attention to finite dimensional
Hausdorff, locally compact $C^{\infty}$-manifolds.\\
We will start by introducing differential operators in
$\mathbb{R}^{n}$, giving an algebraic characterization that will
make us able to deal with differential operators on an arbitrary
manifold (for a mathematical treatment of the topic see \cite{gra}).\\
Let us consider the algebra $\CMcal{A}=C^{\infty}(\mathbb{R}^{n})$
of infinitely differentiable functions on $\mathbb{R}^{n}$: in
standard textbooks (e.g. \cite{vino} ) a \textit{differential
operator of degree at most k} is defined as a linear map
$\Di^{k}:\CMcal{A}\to\CMcal{A}$ of the form:%%
%%citare Vinogradov
\begin{equation}\label{equation:diffop}
\Di^{k}=\sum_{|\sigma|\leq{k}}g_{\sigma}\frac{\partial^{|\sigma|}}{\partial{x_{\sigma}}}\qquad{g_{\sigma}\in\CMcal{A}}
\end{equation}
where $\sigma=(i_{1},\ldots,i_{n})$,
$|\sigma|=i_{1}+i_{2}+\cdots+i_{n}$ and%%
\begin{equation}
\frac{\partial^{|\sigma|}}{\partial{x_{\sigma}}}=
\frac{\partial^{|\sigma|}}{\partial{x_{1}^{i_{1}}\cdots\partial{x_{n}^{i_{n}}}}}
\end{equation}
This is a standard definition; however, one can give an algebraic
characterization suitable for a generalization to arbitrary smooth
manifolds. One can start considering the following commutator
relation%%%
%%%
\begin{equation}
\left[\frac{\partial}{\partial{x_{i}}},\hat{f}\right]=\frac{\partial{f}}{\partial{x_{i}}}
\end{equation}
where $\hat{f}$ is understood to be the multiplicative operator
$\hat{f}:g\to{fg}$, with $f,g\in\CMcal{A}$; then one can verify
that%%
\begin{equation}
\left[\frac{\partial^{|\sigma|}}{\partial{x_{\sigma}}},\hat{f}\right]=\sum_{\tau+\nu=\sigma}c_{\tau}
\frac{\partial^{|\tau|}{f}}{\partial{x_{\tau}}}\frac{\partial^{|\nu|}}{\partial{x_{\nu}}}
\end{equation}
%%%%%%%%%
with $|\tau|>0$ and $c_{\tau}$ constants. From this it easily
follows that%%
\begin{equation}
\left[\Di^{k},\hat{f}\right]=\sum_{|\sigma|\leq{k}}g_{\sigma}
\left[\frac{\partial^{|\sigma|}}{\partial{x_{\sigma}}},\hat{f}\right]
\end{equation}
%%%%%
is a differential operator of degree at most $k-1$. Iterating for
a set of $k+1$ functions $f_{0}, f_{1}, \ldots ,f_{k}\in
\CMcal{A}$, one finds that
\begin{equation}\label{equation:algdiffop}
  [[ \ldots [\Di^{k},\hat{f}_{0}], \hat{f}_{1}], \ldots ,\hat{f}_{k}]=0
\end{equation}
%%%%%%%%
The important fact is that it is possible to prove the converse,
namely that a linear operator satisfying the property above for
each set of $k+1$ elements in $\CMcal{A}$  is
necessarily of the form (\ref{equation:diffop}).\\
Using this algebraic characterization, one can generalize the
notion of differential operator to an arbitrary manifold without
referring to coordinates, by using its algebra of smooth
functions. That is, one takes equation (\ref{equation:algdiffop})
as the definition of a differential operator with degree equal to
the least number of elements of the algebra satisfying this
equation minus 1.\\
In particular, differential operators of degree 1 which annihilate
constant functions within the algebra $\CMcal{A}$ (the constants)
are derivations of $\CMcal{A}$, since it follows from equation
\ref{equation:algdiffop} that they satisfy Leibniz rule with
respect to the Abelian product in $\CMcal{A}$. It is well known
that derivations on the algebra of functions on a manifold are
vector fields on that manifold.\\
Moreover, one can deal with differential operators on a manifold
$\emme$ as the elements of the enveloping algebra of the (infinite
dimensional) Lie algebra of vector fields on $\emme$, with the
associative product obtained from the ``composition'' of derivations,
 and functions over $\emme$. \\
The set of all the differential operators of any degree on
$\emme$, which we denote $\CMcal{D}(\emme)$, can be given a
structure of a graded algebra, and also of a module over the
algebra $\FM$ (see e.g. \cite{vino}).  If the manifold is
parallelizable, i.e. if the cross sections of the tangent bundle
are a free module on $\CMcal{A}$, one has that the differential
operators are a free module on $\CMcal{A}$, with a basis given by
monomials in a finite number of vector fields and the identity
function. If the manifold is not parallelizable we have projective
modules. The case of free modules is simpler and we shall make
some general statements in such a framework. Before doing that,
let us now clarify the notion of projectabily of a differential
operator of an arbitrary
degree, which will be at the core of our following analysis.\\
The preliminary ingredients are two manifolds $\emme$ and $\enne$,
and a submersion\footnote{Obviously this requires
$m=\mathrm{dim}{\emme}>n=\mathrm{dim}{\enne}$.}
$\pi:~\emme\to\enne$. The starting point is the well known
(\cite{godbillon}, \cite{MSSV}) definition of projectability of a
vector field on $\emme$ onto a vector field on $\enne$, i.e.
$\mathbf{X}\in{\chi\left(\emme\right)}$ is projectable with
projection $\mathbf{\tilde{X}}\in{\chi\left(\enne\right)}$  if%%
 \begin{equation}\label{eq:proiet1}
\xymatrix{T\emme \ar[r]^{T\pi}     & T\enne\\
 \emme \ar[u]^{\mathbf{X}} \ar[r]_{\pi}        & \enne
 \ar[u]_{\tilde{\mathbf{X}}}}
\end{equation}
%%
%%%%%%%
Following the previous line, one can recast this definition in
purely algebraic terms, which is more suitable for a
generalization. At this aim, we will use the fact that an
arbitrary manifold $\emme$ is pefectly encoded in its algebra of
smooth functions $\FM=C^{\infty}\left(\emme\right)$. In this
context, one knows by general results that the submersion
$\pi:\emme\to\enne$ gives rise to an injective homomorphism
$\pi^{*}:\FN\to\FM$, that allows one to consider $\FN$ as a
subalgebra of $\FM$.\\
Since a vector field on a manifold  is a derivation of its algebra
of functions, one can prove (\cite{MSSV}) that the above
definition \ref{eq:proiet1} of projectability is equivalent to
require that
%%forse dim
\begin{equation}
  \mathbf{X}( \FN)\subset\FN
\end{equation}
In other words $\mathbf{X}$ is projectable if the subalgebra $\FN$
is invariant under the action of $\mathbf{X}$ as a derivation.
When $\mathbf{X}$ is projectable,  its projection is \,
$\mathbf{\tilde{X}}=\mathbf{X}|_{\FN}$, i.e. the restriction
of $\mathbf{X}$ to $\FN$ as a linear operator. \\
These considerations lead us naturally to an extension of the
definition of projectability for a differential\footnote{Actually
the definition we suggest is valid for a general linear operator,
since it involves only the linear structure of the subalgebra}
operator of arbitrary degree, in that a differential operator
$\Di^{k}$ of degree at most $k$ will be called projectable (with
respect to
$\pi$) iff %
\begin{equation}\label{eq:proiet2}
\Di^{k}(\FN)\subset\FN
\end{equation}
As above, the projection of $\Di^{k}$ will be
$\tilde{\Di}^{k}=\Di^{k}|_{\FN}$.\\[2mm]
\textit{Remark}:\:%subalgebra dell env alg
 The projectable differential operators (with respect to a given
map) are a subalgebra of the whole graded algebra of differential
operators on a given manifold; in fact, it is easy to show that
they are closed under the operations of addition and composition
of linear maps. Anyhow, they cannot be given a structure of
submodule of the module $\CMcal{D}(\CMcal{M})$ over the algebra
$\CMcal{F}(\CMcal{M})$, but only of a module over the algebra
$\CMcal{F}(\CMcal{N})$.
%probabilmente inserire il rapporto col push-forward se lo usiamo dopo
%
\subsection{Differential operators and the KS
fibration}\label{sec:doks} We are now going to give few additional
details about the mathematical setting of our case of study,
underlining its main characteristic
features in relationship with the above setting.\\
In this section we will deal only with differential operators on
$\FM$, not considering the additional structures (e.g. the Hilbert
space structure) that neverthless are necessary for quantum
mechanics; we postpone such consideration to a later section,
pointing out that the following results are valid in the more
general setting of the reduction of arbitrary differential
equations.\\
In the case at hand, we will consider the problem of the
projectability of differential operators on
$\mathbb{R}^{4}-\{0\}\equiv\rqq$ on differential operators on
$\mathbb{R}^{3}-\{0\}\equiv\rt$ with respect to the so-called \ks
map. In particular, we will consider in greater detail second
order differential operators, since the ones we are interested in
are of this form. For this purpose, let us recall
some features of this map, first introduced in \cite{ks}.\\
The main idea behind the construction of this map relies on the
observation that $\rt$ and $\rqq$ may be given a structure of
trivial bundles over spheres, i.e. $\rt=S^{2}\times\mathbb{R}^{+}$
and $\rqq=S^{3}\times\mathbb{R}^{+}$. Then one starts from the
well know Hopf map $\pi_{H}:S^{3}\to{S^{2}}$ (\cite{hopfmap});
identifying $S^{3}$ with $SU(2)$, $\pi_{H}$ may be represented as
(\cite{balamarmo})%%
\begin{equation}
\pi_{H}:s\in{SU(2)}\to{\vec{x}}\in{S^{2}}:s\sigma_{3}s^{-1}=x^{i}\sigma_{i}
\end{equation}
where $\sigma_{i}$ are the Pauli matrices and $x^{i}$ are
cartesian coordinates in $\rt$. Now one may (not uniquely) extend
the Hopf map to $\rqq\to\rt$ by introducing polar coordinates in
$S^{3}\times\mathbb{R}^{+}$; setting%%
\begin{equation}
g=Rs\quad\text{with}\quad{}s\in{SU(2)},\:{R\in\mathbb{R}^{+}}
\end{equation}
we may define%%
\begin{equation}\label{equation:ksmap}
\pi_{KS}:g\in{\rqq}\to\vec{x}\in{\rt}:\:x^{k}\sigma_{k}=g\sigma_{3}g^{\dag}=R^{2}s\sigma_{3}s^{-1}
\end{equation}
In a cartesian system of coordinates one has esplicitly%%
\begin{eqnarray}
x_{1}&=&2(y_{1}y_{3}+y_{2}y_{0})\nonumber\\
x_{2}&=&2(y_{2}y_{3}-y_{1}y_{0})\\
x_{3}&=&y_{1}^{2}+y_{2}^{2}-y_{3}^{2}-y_{0}^{2}\nonumber
\end{eqnarray}
Moreover one finds that%%
\begin{equation}
\sqrt{x^{i}x_{i}}=r=R^{2}=y^{i}y_{i}
\end{equation}
%%%%%%
It is important to notice that the map $\pi_{KS}$ above
constructed, which in the following we refer to as KS-fibration,
defines a principal fibration $\rqq\to\rt$ with structure group
$U(1)$. The compactness of the fiber will be a useful feature when
we will discuss the quantum setting: moreover the very structure
of a fibration allows an easy ``dualization'' of the case at hand.
Infact, being $\pi_{KS}$ a submersion, we can embedd injectively
$\CMcal{F}\left(\rt\right)$ into $\CMcal{F}\left(\rqq\right)$ by%%
\begin{equation}\label{equation:embbed}
\pi_{KS}^{*}:f\in{\CMcal{F}\left(\rt\right)}\to{}f\circ\pi_{KS}\in\CMcal{F}\left(\rqq\right)
\end{equation}
In particular, the above map realizes $\CMcal{F}\left(\rt\right)$
as the subalgebra of $\CMcal{F}\left(\rqq\right)$ made up by
functions which are constant along the fibers.\\
In this algebraic context, we may now investigate the
projectability of differential operators from $\rqq$ onto $\rt$
and in particular of second order ones: at this aim, it will be
very useful to characterize $\CMcal{F}\left(\rt\right)$ using
vector fields defined on $\rqq$, in the following way.\\
From equation (\ref{equation:ksmap}), one may observe that the
orbits of the one parameter group $\exp(i\lambda\sigma_{3})$
acting by right multiplication on $S^{3}\times{\mathbb{R}^{+}}$
are the fibers of the KS-fibration. Hence, there is a natural
action by automorphisms of $\exp(i\lambda\sigma_{3})$ on
$\CMcal{F}\left(\rqq\right)$ given by%%
\begin{equation}
(U_{\lambda}f)(g)=f(g\exp(i\lambda\sigma_{3}))
\end{equation}
with $g=Rs,\:s\in{SU(2)},\:{R\in\mathbb{R}^{+}}$. Now, by equation
(\ref{equation:embbed}), the elements $f$ of
$\CMcal{F}\left(\rt\right)$ are the only ones which fulfill the
condition%%
\begin{equation}
U_{\lambda}f=f\quad\forall\:\lambda\in\mathbb{R}
\end{equation}
This condition can be espressed by saying that the infinitesimal
generator $\mathbf{X}_{3}$ of $U_{\lambda}$ annihilates
\textit{all and only} the functions costant along the fibers, i.e.
$\text{Ker}\mathbf{X}_{3}=\CMcal{F}\left(\rt\right)$. Moreover,
$\mathbf{X}_{3}$ is a left invariant vector field for the action
of $SU(2)$, and in cartesian coordinates it has the expression%%
\begin{equation}\label{X3}
\mathbf{X}_{3}=y^{0}\frac{\partial}{\partial{y^{3}}}-y^{3}\frac{\partial}{\partial{y^{0}}}+y^{1}\frac{\partial}{\partial{y^{2}}}-y^{2}\frac{\partial}{\partial{y^{1}}}
\end{equation}
By means of the above vector field, we may give a  condition for
the projectability of differential operators of any degree. Let us
consider the algebra $\CMcal{C}$ generated by the monomials in
$\mathbf{X}_{3}$ of any degree and $\Frt$ (i.e. a central Lie
algebra extension of the enveloping algebra of $\mathbf{X}_{3}$).
Projectable operators with respect to $\pi_{KS}$ are then given by
the \textit{normalizer} of this extension in the algebra
$\CMcal{D}(\rqq)$ of differential operators on $\Frqq$, i.e.
the set of elements $\Di^{N}$ of $\CMcal{D}(\rqq)$ such that%
\begin{equation}\label{eq:normalizer}
  [\Di^{N}, \mathrm{C}]\in\CMcal{C} , \qquad \forall\,\mathrm{C}\in\CMcal{C}
\end{equation}
One can easily show that if $\Di^{N}$ satisfies the above
equation, then it leaves invariant the subalgebra $\Frt$; indeed,
one can prove that the commutator $[\Di, \mathbf{X}_{3}]$ is
necessarily an \textit{homogeneous} element of $\CMcal{C}$. It
follows that
\begin{equation}
  \left[\Di^{N},\mathbf{X}_{3}\right](f)=0\qquad\forall\,f\in\Frt
\end{equation}
which implies
\begin{equation}
  \mathbf{X}_{3}(\Di^{N}f)=0 \qquad \forall\,f\in \Frt
\end{equation}
i.e. $\Di^{N}$ leaves the subalgebra $\Frt$ invariant, that is it
is projectable according to our definition (\ref{eq:proiet2}).\\
The converse result is quite immediate, being aware of the fact
that adding to a projectable differential operator an expression
in $\mathbf{X}_{3}$ with arbitrary coefficients in $\Frqq$ does
not alter its projectability property.\\
The above characterization is general; however, for simplicity it
is often useful to deal with the centralizer $\CMcal{D}^{C}$ of
$\mathbf{X}_{3}$ (instead of the normalizer of $\CMcal{C}$), i.e.
the set of differential operators in $\CMcal{D}(\rqq)$ that
commutes with $\mathbf{X}_{3}$%
\begin{equation}\label{equation:comm}
\left[\Di^{k},\mathbf{X}_{3}\right]=0
\end{equation} %
Differential operators in the centralizer are obviously
projectable, since they satisfy eq. \ref{eq:normalizer} in a
trivial way. Although this is not a general condition, it will be
very useful in our study, since one can show that, in the case at
hand, the projections of all the elements in the centralizer of
$\mathbf{X}_{3}$ cover all the differential operators in $\rt$.
Because  this property is sufficient for our future purposes, in
the following we will deal only with differential operators in
this centralizer; thus, we will caracherize $\CMcal{D}^{C}$ in some more detail. \\
The centralizer $\CMcal{D}^{C}$ forms a subalgebra of
$\CMcal{D}(\rqq)$; in particular it is a module over the
subalgebra $\Frt$, not over the whole algebra because the
pointwise product of a function which is \textit{not} costant
along the fibers with an element in $\Frt$ does not give again a
function costant along the fibers. The interesting result is that
this subalgebra $\CMcal{D}^{C}$ may be constructed in the
following way from a set of projectable fields and functions which
are constant along the fibers. One starts by introducing a
``basis'' of $\chi\left(~S^{3}\times\mathbb{R}^{+}\right)$ set up
by three basis vector fields on $S^{3}$ (as it is parallelizable)
and a field on $\mathbb{R}^{+}$, which commutes with all the
others, because of the structure of Cartesian product.\\
One can choose a basis on $S^{3}$ given by the three right
invariant vector field, $\mathbf{Y}_{1}$, $\mathbf{Y}_{2}$,
$\mathbf{Y}_{3}$, that, from general theory of Lie groups, commute
with $\mathbf{X}_{3}$, since it is a left invariant vector field.
Adding a vector field $\mathbf{R}$ on $\mathbb{R}^{+}$, we have a
basis of $\chi\left(\rqq\right)$ formed by
projectable vector fields.\\
Having in mind what stated in the previous section, the
differential operators on $\Frqq$ belong to the algebra generated
by monomials in
$\mathbf{Y}_{1},\mathbf{Y}_{2},\mathbf{Y}_{3},\mathbf{R}$ and
elements in $\Frqq$. In particular, the above monomials are
projectable linear maps, in that their commutator with
$\mathbf{X}_{3}$ satisfy condition (\ref{equation:comm}), as can
easily be seen just using the
Leibniz rule with respect to the composition.\\
Now, consider the associative algebra %$\CMcal{D}^{P}$
generated by
monomials in the above basis fields and elements of $\Frt$.
Because the relation%%
\begin{equation}
\left[fD,\mathbf{X}_{3}\right]=f[D,\mathbf{X}_{3}]-\mathbf{X}_{3}(f)D
\end{equation}
is satisfied by an arbitrary function $f$ and a linear map $D$,
all the elements of the above algebra %$\CMcal{D}^{P}$
commute with $\mathbf{X}_{3}$, i.e. belong to $\CMcal{D}^{C}$,
 and so are projectable.\\
By the converse, decomposing (uniquely) a differential operator
$\Di$ in the projectable basis, if it belongs to $\CMcal{D}^{C}$
condition (\ref{equation:comm}) immediately implies that the
``functional coefficients'' of $\Di$ are
annihilated by $\mathbf{X}_{3}$.\\
So we have given a useful characterization of the centralizer of
$\mathbf{X}_{3}$ as the elements generated by monomials in the
above basis with \A\A functional coefficient'' belonging to
$\Frt$. Using it we are able to show that the projections of the
elements in $\CMcal{D}^{C}$ actually cover the differential
operators on
$\rt$.  \\
One can start from considerations about the submodule of the
vector fields. Since, when a vector field is projectable, the
push-forward of this vector field evaluated at each point defines
again a vector field, which coincides with its projection, one has
an operative way to obtain the projection of a vector field. In
this way one easily find that the push-forward of the vector
fields of the basis $(\mathbf{R}, \mathbf{Y}_{1}, \mathbf{Y}_{2},
\mathbf{Y}_{3})$ of $\chi(\rqq)$ with respect to the KS map are
respectively a vector field along $r$ and the generators of the
three rotation of $\rt$.\\
Thus, the module $M$ generated by the above basis, over the
algebra $\Frt$ projects onto the module of the vector fields on
$\rt$ over the algebra $\Frt$: in fact, a basis of the first
projects onto a system of generators of the second, and the
map is $\Frt$-linear, the underlying algebra being the same.\\
Using the fact that monomials in the elements of a system of
generators (resp. a basis) of the Lie algebra of the vector fields
on $\CMcal{M}$ provide a systems of generator (resp. a basis) for
the corresponding enveloping algebra, one can also show that the
module of differential operators on $\rqq$ over the algebra
$\CMcal{F}(\rt)$ projects onto the module of differential
operators on $\rt$.
%%%
%%
\section{Reduction and unfolding: quantum aspects}
In this section we will use the tools introduced so far to study
the problem of reduction of dynamical systems in quantum
mechanics: actually, we will be concerned with the inverse
procedure, the unfolding, which is highly not unique. However, we
will focus on the quantum Kepler problem, i.e. the hydrogen atom,
having in mind to clarify how an unfolding procedure may be
defined for a quantum system. We leave aside the general treatment
for a future study.\\
The link with the previous section relies on the fact that a
quantum dynamical system with Hamiltonian operator
$\hat{\mathrm{H}}$ is supposed to evolve in time accordingly to
the Schr\"{o}edinger equation%%
\begin{equation}
i\hbar\dot{\psi}(t)=\hat{\mathrm{H}}\psi(t)
\end{equation}
where $\psi$ is a vector in a ``functional'' Hilbert space
$\CMcal{H}_{\CMcal{N}}$ constructed over a ``configuration
manifold'' $\enne$. Generally, $\hat{\mathrm{H}}$ is represented
on $\CMcal{H}_{\enne}$ in terms of a (non homogeneous)
differential operator of second degree: hence, it is natural to
define an unfolding procedure for a quantum system in the
following way. First of all, one searches for a higher dimensional
configuration manifold $\emme$ with a submersion
$\pi:\emme\to\enne$, and a differential operator
$\hat{\mathrm{H}}^{'}$ which projects onto $\hat{\mathrm{H}}$ with
respect to $\pi$. Then, one constructs an Hilbert space
$\CMcal{H}_{\emme}$ of ``wave functions'' which contains
$\CMcal{H}_{\enne}$ as a subspace, and on which the operator
$\hat{\mathrm{H}}^{'}$ extends the operator $\hat{\mathrm{H}}$. As
stated above, all the steps in this procedure are not unique.
However, one may restrict the arbitrariness by the means of
physical motivations, e.g. by introducing symmetries, etc:
this is just what happens in our case of study.\\
Let us consider the following differential operator%%
\begin{equation}
\hat{\mathrm{H}}=-\frac{\Delta_{3}}{2}-\frac{k}{r}
\end{equation}
where $\Delta_{3}$ is the Laplacian operator on $\rt$ associated
with the Euclidean metric, $r$ is the radial coordinate, $k>0$ is
a coupling constant, and the Hilbert space $\CMcal{H}_{\rt}$ is
$\CMcal{L}^{2}\left(\rt,d^{3}x\right)$. It is well known that
$\hat{\mathrm{H}}$ describes the hydrogen atom in the center of
mass system, where we have set $m=\hbar=1$, and and is the quantum
analogue of the Kepler problem. It seems thus natural to us to
relate the hydrogen atom with a differential (inhomogeneous)
operator on $\rqq$ projectable with respect to the KS-fibration,
as it happens in the classical counterpart. So, we search for a
second order non-homogeneous differential operator
$\hat{\mathrm{H}}'$ on $\rqq$ which projects onto
$\hat{\mathrm{H}}$; since the multiplicative operator
$k\hat{R}^{-2}$ obviously project on $k\hat{r}^{-1}$ (since
$\pi^{*}_{KS}(r^{-1})=R^{-2}$), we are left with the search for a
second order differential  operator $D$ that projects on the
Laplacian $\Delta_{3}$. It is obvious that there may be many such
operators; however, we may drastically reduce this arbitrariness
by symmetry requirements. Indeed, one knows that the algebra of
invariance of the hydrogen atom in three dimensions is
$\mathfrak{so}(4)\approx\mathfrak{su}(2)\times\mathfrak{su}(2)$,
generated by the angular momentum and the Runge-Lenz vector
(\cite{Wyb},\cite{englefield}). It seems therefore reasonable to
ask that the unfolding quantum system in 4 dimensions, and so
$\mathrm{D}$, shares the same symmetry property, so that
$\mathfrak{su}(2)\oplus\mathfrak{su}(2)$ is at least a subalgebra
of its algebra of invariance.\\
This request of invariance can be stated in term of invariance
with respect to the algebra
$\mathfrak{su}(2)\oplus\mathfrak{su}(2)$ generated by the direct
product of the left action and the right action of
$\mathrm{SU(2)}$ that one naturally has on $S^{3}\approx
\mathrm{SU(2)}$. To exploit this requirement, it is useful to
decompose the problem in a spherical and a radial part (e.g. using
the basis introduced in section \ref{sec:doks}). The radial part
is obviously invariant under this action, so we only have to
impose that the part on $S^{3}$ is invariant under left and right
action of $SU(2)$. One knows (\cite{helgason}) that the only
second order differential operator with the above property is
$\Delta^{S}_{3}$, the Laplacian associated with $S^{3}$. This
excludes mixed terms (composition of operators along $S^{3}$ with
the one along $\mathbb{R}$), implying that $\mathrm{D}$
\textit{must} be of the form%%
\begin{equation}\label{eq:oppr}
\mathrm{D}=f(R)\frac{\partial^{2}}{\partial{R}^{2}}+g(R)\frac{\partial}{\partial{R}}+h(R)\Delta^{S}_{3}+c(R)
\end{equation}
where $R$ is the radial coordinate in $\rqq$.\\
Before imposing that $\mathrm{D}$ project on $\Delta_{3}$, we
recall that $\Delta_{3}$ can be expressed as%%
\begin{equation}\label{eq:delta3}
\Delta_{3}=\frac{\partial^{2}}{\partial{r}^{2}}+\frac{2}{r}\frac{\partial}{\partial{r}}
+\frac{1}{r^{2}}\Delta^{S}_{2}
\end{equation}
where $\Delta^{S}_{2}$ is the Laplacian associated with $S^{2}$.\\
Moreover, it can be easily proven that $\Delta^{S}_{3}$ does
project onto $\Delta^{S}_{2}$ with respect to the Hopf
fibration.\\
So, the requirement that the differential operator $\mathrm{D}$ as
in eq. \ref{eq:oppr} projects onto $\Delta_{3}$ as in eq.
(\ref{eq:delta3}) with respect to the KS-fibration becomes just a
conditions on the functions $f,g,h,c$, that in this way are
fixed.\\
In the end, one finds that the differential operator that projects
on $\Delta_{3}$, with the additional condition of invariance under
$\mathfrak{su}(2)\oplus\mathfrak{su}(2)$
is:%
\begin{equation}
\mathrm{D}=\frac{1}{4R^{2}}\frac{\partial^{2}}{\partial{R}^{2}}+
\frac{1}{2R^{3}}\frac{\partial}{\partial{R}}+\frac{1}{4R^{4}}\Delta^{S}_{3}
=\frac{1}{4R^{2}}\Delta_{4}
\end{equation}%
where $\Delta_{4}$ is the Laplacian operator associated with the
Euclidean metric.\\
It follows that $\hat{\mathrm{H}}^{'}$ is expressed as%%
\begin{equation}
\hat{\mathrm{H}}^{'}=-\frac{1}{2}\frac{1}{4R^{2}}\Delta_{4}-\frac{k}{R^{2}}
\end{equation}
Usually, the operator $\hat{\mathrm{H}}^{'}$ is referred
to as the \textit{conformal Kepler} Hamiltonian.\\
The last step in this unfolding procedure involves the costruction
of a Hilbert space $\CMcal{H}_{\rqq}$ of functions over $\rqq$ on
which $\hat{\mathrm{H}}^{'}$ is defined. At first sight one may
think to choose $\mathcal{L}^{2}\left(\rqq,\de^{4}y\right)$ with
$\de^{4}y$ the Lebesgue measure on $\mathbb{R}^{4}$: however, the
operator $\hat{\mathrm{H}}^{'}$ is by no means symmetric on this
Hilbert space. Hence, we may search for a measure on $\rqq$ which
induces a scalar product on $\Frqq$ so that $\hat{\mathrm{H}}^{'}$
is symmetrically defined. Actually, the operator
$\hat{\mathrm{H}}^{'}$ should be (essentially) selfadjoint on its
domain in order to assure the existence of a unitary dynamics,
i.e. a one parameter group of a unitary time evolution: however,
we leave for a later section the analysis of this question, in
which we will also argue that ``the search for a selfadjoint
Hamiltonian'' is an echo of the reparametrization of
the dynamical field which happens in the classical case.\\
Let us consider the Hilbert space
$\CMcal{H}_{\rqq}=\CMcal{L}^{2}\left(\rqq,4R^{2}\de^{4}y\right)$:
if one chooses the set of $C^{\infty}$ functions with compact
support%%
\begin{equation}
\CMcal{D}\equiv{}C^{\infty}_{0}\left(\rqq\right)
\end{equation}
which clearly belongs to $\CMcal{H}_{\rqq}$, the operator
$\hat{\mathrm{H}}^{'}$ becomes Hermitian on $\CMcal{D}$. Moreover,
$\CMcal{D}$ is dense in $\CMcal{H}_{\rqq}$, as one can prove by
using the unitary operator%%
\begin{equation}
U:\CMcal{L}^{2}\left(\rqq,4R^{2}\de^{4}y\right)\to\CMcal{L}^{2}\left(\rqq,\de^{4}y\right):\varphi\to{2R}\varphi
\end{equation}
which leaves the set $\CMcal{D}$ invariant. Then,
$\hat{\mathrm{H}}^{'}$ is symmetric on $\CMcal{D}$: actually, one
can prove that $\hat{\mathrm{H}}^{'}$ is essentially selfadjoint
on $\CMcal{D}$, by using the same arguments of \cite{RS2} as in
the proof of the essential self-adjointness of Laplacian operators
in $\mathbb{R}^{n}$.\\
Now we will show that $\CMcal{H}_{\rt}$ is contained as a subspace
in $\CMcal{H}_{\rqq}$: this may be done easily using the following
geometrical argumentation. The measure $\de^{3}x$ and
$4R^{2}\de^{4}y$ are associated with two volume forms, which we
denote by $\boldsymbol{\mathbf{\mu}_{3}}$ and
$\boldsymbol{\mathbf{\mu}_{4}}$ respectively. Explicitely, in a
Cartesian system of coordinate, one has%%
\begin{eqnarray}
\boldsymbol{\mu}_{3}&=&\de x_{1}\wedge{\de x_{2}}\wedge{\de x_{3}}\nonumber\\
\boldsymbol{\mu}_{4}&=&4R^{2}\de y_{0} \wedge{\de y_{1}}
\wedge{\de y_{2}} \wedge{\de y_{3}}
\end{eqnarray}
Now, the following relation holds between $\boldsymbol{\mu}_{3}$
and $\boldsymbol{\mu}_{4}$%%
\begin{equation}
i_{\mathbf{X}_{3}}(\boldsymbol{\mu}_{4})=\pi_{KS}^{*}(\boldsymbol{\mu}_{3})
\end{equation}
and then we may factorize $\boldsymbol{\mu}_{4}$ in the form%%
\begin{equation}
\boldsymbol{\mu}_{4}=\pi_{KS}^{*}(\boldsymbol{\mu}_{3})\wedge{\boldsymbol{\Theta}}_{3},
\end{equation}
where $\boldsymbol{\Theta}_{3}$ is a dual field\footnote{Any dual
fields of form suits well for our purposes, and we may select one
 e.g. by requiring left invariance} to $\mathbf{X}_{3}$.\\
Now consider two functions $\tilde{f},\tilde{g}$ costant along the
fibers: from the previous section we know that such functions are
of the form $\tilde{f}=\pi^{*}_{KS}(f)$ and
$\tilde{g}~=~\pi^{*}_{KS}(g)$, with $f$ and $g$ uniquely
determined. Then, by using some general theorems (\cite{matone},
\cite{greub}) about the integration of forms along fibers of the
bundle space of a fibration, one has that%%
\begin{eqnarray}
\int_{\rqq}\bar{\tilde{f}}\,\tilde{g}\,\boldsymbol{\mu}_{4}&=&\int_{\rqq}\bar{\tilde{f}}\,\tilde{g}\,\pi_{KS}^{*}\,(\boldsymbol{\mu}_{3})\wedge{\boldsymbol{\Theta}}_{3}\nonumber\\
&=&\int_{\rt}\bar{f}\,g\,\boldsymbol{\mu}_{3}\int_{U(1)}\boldsymbol{\Theta}_{3}\nonumber\\
&=&c\:\int_{\rt}\bar{f}\,g\,\boldsymbol{\mu}_{3}
\end{eqnarray}
where $c$ does not depend on the pair of functions $\tilde{f}$ and
$\tilde{g}$. To obtain the result above, we have mainly used that
$\boldsymbol{\Theta}_{3}$ is dual to the tangent field to the
fiber, and that the fiber itself is compact. So, we have been able
to ``integrate out'' the fiber contribute, hence to show that the
KS-map $\pi_{KS}^{*}$ gives rise (up to a constant scaling factor)
to an isometric embedding $U_{KS}$ of $\CMcal{H}_{\rt}$ into
$\CMcal{H}_{\rqq}$.\\%%
\subsection{Eigenvalue equation}
At this point we may study the eigenvalue equation for
$\hat{\mathrm{H}}^{'}$ on $\CMcal{H}_{\rqq}$: obviously, because
of our unfolding procedure, the set of solutions of this equation
contains properly that of the eigenvalue problem of the hydrogen
atom on $\CMcal{H}_{\rt}$, and are precisely those solutions which
are constant along the fibers.\\ %So, we can obtain all the
%eigenfunctions of $\hat{H}$ (and the corresponding eigenvalues) by
%solving the eigenvalue equation for $\hat{\mathrm{H}}^{'}$ on
%$\CMcal{H}_{\rqq}$ and then selecting those
%which are constants along the fibers.  \\
Then, we consider the following equation%%
\begin{equation}\label{eq:hydrogen}
\left(-\frac{1}{2}\frac{1}{4R^{2}}\Delta_{4}-\frac{k}{R^{2}}\right)\psi-E\psi=0
\end{equation}
which defines a subspace in $\Frqq$.\\
This same subspace is also defined by the equation
\begin{equation}\label{harmonic}
\left(-\frac{1}{2}\Delta_{4}-{4R^{2}}E\right)\psi-4k\psi=0
\end{equation}
So, if we are interested in bound states of the hydrogen atom
($E_{n}<0$), the equation (\ref{eq:hydrogen}) may be conveniently
replaced by equation (\ref{harmonic}), which represents an
eigenvalue equation for a ``family of isotropic quantum harmonic
oscillator'' with frequence $\omega(E)=\sqrt{-8E}$ depending on
the energy. We then find a ``metamorphosis'' (\cite{Hietarinta})
of the coupling costant $k$ into an eigenvalue; we can determine
the ``admissible frequencis'' by imposing that%%
\begin{equation}
4k=\omega(E)(N+2),\qquad{}N\in\mathbb{N}
\end{equation}
and then%%
\begin{equation}
E_{N}=-\frac{2k^{2}}{(N+2)^{2}}
\end{equation}%%
As we stated above, not all the solutions of equation
(\ref{harmonic}) are solutions of the eigenvalue problem for the
three-dimensional hydrogen atom: infact, we have to select the
projectable ones, i.e. the functions $\psi$ which are constant
along the fibers. So, we must impose%%
\begin{equation}
\mathbf{X}_{3}\psi_{N}=0
\end{equation}
Recalling the expression (\ref{X3}), and using the knowledge of
the solutions of equation (\ref{harmonic}) (products of Hermite
functions), %after a lenghty calculation
one finds that $N$ should
be an even natural number %%
\begin{equation}
N=2m,\qquad{}m\in\mathbb{N}
\end{equation}
Then, the eigenvalues corresponding to the projectable
eigenfunctions are%%
\begin{equation}
E_{m}=-\frac{k^{2}}{2(m+1)^{2}},\qquad{}m\in\mathbb{N}
\end{equation}%%
and one easily recognizes the energy level of the hydrogen atom,
taking into account the fact that $m$ starts from 0.\\
Thus, we have found the spectrum of the hydrogen atom and the
corresponding eigenfunctions (with the correct multiplicity) by
solving the unfolding system, which is related to a family of
harmonic oscillators.
%%%
%
%
\subsection{Reduction of symmetry}
For the sake of completeness, in this section we briefly discuss
how the symmetry algebra of the conformal Kepler problem in 4
dimensions reduces to the symmetry algebra of the Kepler problem
in 3 dimensions. This topic has been the subject of extensive
studies, and here we just want to present it in our perspective,
i.e. in the setting of the projectability of differential
operators on a manifold; we refer to the literature for details.\\
First of all, one has to carachterize the symmetry algebra of the
conformal Kepler problem. This can be done, and was done (see
\cite{iwaiq}), starting from the symmetry of the harmonic
oscillator: one first restricts to each eigenspace, where the
conformal Kepler problem is related to an harmonic oscillator, and
then extends the algebra so obtained to (a suitable domain of) the
whole Hilbert space (cfr. with the classical case, ref.
\cite{ruclass}). It is well known that the symmetry algebra of an
n-dimensional harmonic oscillator is $\mathfrak{u}(n)$, whose
generators are, in terms of differential operators\footnote{Their
expression in terms of the creation and annihilation operators is
more familiar, but not suitable for their extension
from each eigenspace to the whole Hilbert space} on $\rqq$\\
\begin{align}
  \hat{\mathrm{L}}_{\alpha\beta}=&y^{\alpha}\frac{\partial}{\partial
  y^{\beta}}-y^{\beta}\frac{\partial}{\partial
  y^{\alpha}}\\
  \hat{\mathrm{D}}^{E}_{\alpha\beta}=&\frac{1}{2}\left(\omega^{2}(E)\,
  y^{\alpha}y^{\beta}+\frac{\partial^{2}}{\partial y^{\alpha}
  \partial y^{\beta}}\right)
\end{align}
There is an isomorphism between this algebra and the algebra of
complex $4\times 4$ anti-Hermitian matrices; each matrix $C$ of
this kind can be split in the sum of two real matrices as $C=A+iB;
\: A=-A^{T}, \: B=B^{T}$: the $\hat{L}_{\alpha\beta}$ correspond
to the antisymmetric ones, the $\hat{D}_{\alpha\beta}$ to the
symmetric ones.
\\
Thus, the operators above, when restricted to each eigenspace of
$\hat{\mathrm{H}}'$ corresponding to the eigenvalue $E$, represent
symmetries for $\hat{H}'$. They can be extended to the whole
Hilbert space, with some care due to the fact that the resulting
operators have to be (essentially) selfadjoint and to commute with
$\hat{H}'$ (on the intersection of their domains). \\
The $\hat{\mathrm{L}}_{\alpha\beta}$ satisfy automatically these
properties (being the generators of the rotations); so do also the
$\hat{\mathrm{D}}_{\alpha\beta}$, providing that one replace the
value of the eigenvalue with the Hamiltonian (for a more careful
explanation see \cite{iwaiq}). So, the symmetry algebra of the
conformal Kepler problem is $\mathfrak{u}(n)$, represented by:
\begin{align}
  \hat{L}_{\alpha\beta}=&y^{\alpha}\frac{\partial}{\partial
  y^{\beta}}-y^{\beta}\frac{\partial}{\partial y^{\alpha}}\\
  \hat{D}_{\alpha\beta}=&\frac{1}{2}\left(-2y^{\alpha}y^{\beta}\hat{\mathrm{H}}'
  +\frac{\partial^{2}}{\partial y^{\alpha}  \partial y^{\beta}}\right)=
  \frac{1}{2}\left( \frac{ y^{\alpha}y^{\beta}}{R^{2}}\Delta_{4}
  + 2k\frac{ y^{\alpha}y^{\beta}}{R^{2}}
  +\frac{\partial^{2}}{\partial y^{\alpha} \partial y^{\beta}}\right)
\end{align}
In our scheme, to obtain the symmetry algebra of the three
dimensional Kepler problem, one has to consider the subalgebra of
the whole symmetry algebra of the unfolding system made of
projectable operators. Since, in our specific case, the
centralizer of $\mathbf{X}_{3}$ covers the differential operators
on $\rt$, we can restrict our analysis to the subalgebra of the
symmetry algebra $\mathfrak{u}(4)$ of differential operators in
this centralizer.\\
To obtain the combinations of $\hat{\mathrm{L}}_{\alpha\beta}$ and
$\hat{\mathrm{D}}_{\alpha\beta}$ that commute with
$\mathbf{X}_{3}$, it is easier to first restrict to each
eigenspace of $\hat{\mathrm{H}}'$, where we can use the
isomorphism with the algebra of matrices, as above. The problem is
then to find those matrices, among the symmetric and antisymmetric
ones, that commute with the matrix representative of
$\mathbf{X}_{3}$: this problem is exactly the same as in the
classical case, and has already been solved in that setting (see
\cite{ruclass}, \cite{iwaioa81}). We will not do it again here,
but give the result in terms of differential operators.\\
As for the antysimmetric ones, we are left with the three
operators $\hat{\mathrm{L}}_{i}$, i.e. the three right invariant
vector fields on the 3-sphere \\
\begin{align}
  \hat{\mathrm{L}}_{1}:=\hat{\mathrm{L}}_{10}+\hat{\mathrm{L}}_{32}=\mathbf{Y}_{1}\nonumber\\
  \hat{\mathrm{L}}_{2}:=\hat{\mathrm{L}}_{02}+\hat{\mathrm{L}}_{13}=\mathbf{Y}_{2}\nonumber\\
  \hat{\mathrm{L}}_{3}:=\hat{\mathrm{L}}_{03}+\hat{\mathrm{L}}_{21}=\mathbf{Y}_{3}
\end{align}\\
As for the $\hat{\mathrm{D}}_{\alpha\beta}$, one obtains the
following combinations:
\begin{align}
  \hat{\mathrm{D}}_{1}&=\frac{1}{2}\left\{\left(y^{1}y^{3}+y^{2}y^{0}\right)(-2E)E + \frac{\partial^{2}}{\partial
  y^{1}\partial y^{3}}+\frac{\partial^{2}}{\partial
  y^{2}\partial y^{0}}\right\}\\
  \hat{\mathrm{D}}_{2}&=\frac{1}{2}\left\{\left(y^{2}y^{3}-y^{1}y^{0}\right)(-2E)+ \frac{\partial^{2}}{\partial
  y^{2}\partial y^{3}}-\frac{\partial^{2}}{\partial
  y^{1}\partial y^{0}}\right\}\\
  \hat{\mathrm{D}}_{3}&=\frac{1}{4}\bigg\{\big((y^{1})^{2}+(y^{2})^{2}-(y^{3})^{2}-(y^{0})^{2}\big)(-2E)+\\
    \quad & \qquad +\frac{\partial^{2}}{\partial (y^{1})^{2}}+\frac{\partial^{3}}{\partial
      (y^{2})^{2}}-\frac{\partial^{2}}{\partial (y^{3})^{2}}-\frac{\partial^{2}}{\partial
        (y^{0})^{2}}\bigg\}
\end{align}
While the antisymmetric operators are well defined on the whole
Hilbert space, for the symmetric one we replace $E$ with the
Hamiltonian operator, as explained above. In the end, we have a
set of 6 differential operators that commute with $\mathbf{X}_{3}$
and satisfy the following commutation relations:
\begin{flalign}
[\hat{\mathrm{L}}_{i},\hat{\mathrm{L}}_{j}]&=i\epsilon_{ijk}\hat{\mathrm{L}}_{k} \nonumber\\
[\hat{\mathrm{L}}_{i},\hat{\mathrm{D}}_{j}]&=i\epsilon_{ijk}\hat{\mathrm{D}}_{k}\nonumber\\
[\hat{\mathrm{D}}_{i},\hat{\mathrm{D}}_{j}]&=i\epsilon_{ijk}\hat{\mathrm{L}}_{k}(-2\hat{\mathrm{H}}')
\end{flalign}%
i.e. they close an $\mathfrak{so}(4)$ algebra\footnote{Obviously,
one can rescale by an appropriate function of the Hamiltonian,
taking into account that we are restricting to the negative part
of its spectrum.}, as happened in the classical case.\\
Now it is quite easy to find the explicit expressions of the
projections of the above operators on
$\CMcal{L}^{2}(\mathbb{R}^{3},\de^{3}x)$, following the method
suggested in the first section. As for the $\hat{\mathrm{L}}_{i}$,
they are the three right invariant vector fields on the 3-sphere,
and we have already pointed out (at the end of section 2.1) that
they project on the three generators of the rotation in $\rt$, so
on the angular momentum
operators on $\CMcal{L}^{2}(\mathbb{R}^{3},\de^{3}x)$.\\
One can also explicitely find the projections of the
$\mathrm{\hat{D}}_{i}$, using that $\hat{\mathrm{H}}'$ projects on
$\mathrm{\hat{H}}'$, the monomials in the $y^{\alpha}$ on the
$x^{i}$ and that
\begin{equation}
  \frac{\widetilde{\partial^{2}}}{\partial y^{\alpha} \partial
  y^{\beta}} =
  \frac{\partial^{2} x^{i}}{\partial y^{\alpha} \partial
  y^{\beta}}\frac{\partial}{\partial x^{i}}+
  \frac{\partial x^{i}}{\partial y^{\alpha}}\frac{\partial x^{j}}{\partial y^{\beta}}
  \frac{\partial^{2}}{\partial x^{i} \partial x^{j}}
\end{equation}
Putting all this together, one finds that the
$\mathrm{\hat{D}}_{i}$ project on the components of the Runge-Lenz
vector on $\CMcal{L}^{2}(\mathbb{R}^{3},\de^{3}x)$.\\
Summarizing, we have found the subalgebra of the whole symmetry
algebra of the unfolding system of projectable operators, and
found the explicit form of the projected operators on
$\CMcal{L}^{2}(\mathbb{R}^{3},\de^{3}x)$, without the need to
refer to a special coordinate system (as in previous works); the
resulting algebra is $\mathfrak{so}(4)$ and the projected
operators are the angular momentun and the Runge-Lenz vector, as
we expected.
%%%
%%%%%
\section{Reparametrization in the quantum realm: a proposal}
In the previous section we have extablished the relationship
between the three-dimensional hydrogen atom and a family of
harmonic oscillators, in the sense that we have linked
\textit{each} eigenvalue equation for the hydrogen atom with
\textit{an} eigenvalue equation for an isotropic harmonic
oscillator with frequency $\omega(E)=\sqrt{-8E}$ depending on the
energy, where the eigenvalue is fixed and equal to the coupling
constan $k$. This is exactly what happens in the classical case
(see \cite{ruclass}): the unfolding of the Kepler problem in three
dimension is achieved via a conformal Kepler problem in 4
dimension, that is related to a family of harmonic oscillators
with frequency depending on the energy of the Kepler problem. We
notice that in the classical case a reparametrization was
involved; thus, to complete the parallelism, we expect that some
kind of reparametrization is involved also in the quantum case.
This section is devoted to this aim, and before doing that we need
a the clarification of \textit{what is} reparametrization in the
quantum realm, since this notion itself is not obviously clear.\\
First of all, we remark that reparametrization, in the classical
setting, is connected with two different aspects, that is
worthwile to clarify separately, in order to better understand
how one can think of it within the quantum setting. \\
The first aspect, the most immediate one, is that the
reparametrization of a vector field preserves its integral curves:
a vector field\, $\mathbf{X}$ and any reparametrization
$\mathbf{\tilde{X}}=f\cdot\mathbf{X}$ have the same integral
curves, but with different parametrizations. In an intuitive way,
when $f=a$ constant, the tangent vector to each integral curve at
each point is dilated by a constant factor $a$, that corresponds
to  changing the parameter of the integral curves from $t$ to
$\tau = a \, t$; this interpratation can be carried over also when
$f$ is a function on the manifold.  In this sense, since the
parameter of the integral curves of the dinamical vector field is
interpreted as the \A\A time'', reparametrization is connected to
a \A\A change of time'', i.e. one changes the way of measuring the
time, dilating it by a factor f, either constant or depending on
the point.\\
The other aspect of reparametrization is related to the
completeness of vector fields. A vector field
$\mathbf{X}\in\chi(\emme)$ is said to be \textit{complete} when
its flow defines an action of the group $\mathbb{R}$ on $\emme$
(of which $\mathbf{X}$ is the infinitesimal generator). A well
known theorem assures that, given a non complete vector
$\mathbf{X}$ on a paracompact manifold $\emme$, it is always
possible to find a (strictly positive) function
$f\in\CMcal{F}(\emme)$ such that
$\mathbf{\tilde{X}}=f\cdot\mathbf{X}$ is complete. In other words,
the reparametrization can be seen as a way to obtain a
complete vector field out of one which is  not complete.\\
Let us now examine what are the possible \A\A quantum''
counterparts of these two aspects. In fact, classically they are
very closely related, being two different ways to see the same
operation, i.e. the multiplication of a vector field by a
function, while in the quantum setting things are not so
straight.\\
As for the first aspect, in the following consideration we
restrict ourselves to maximally superintegrable system, i.e.
$n$-dimensional systems with $2n-1$ constants of the motion
$F_{i}$; we recall that the Kepler problem, as well as the
harmonic oscillator, have this property. In this case (with the
necessary regularity requirements), each integral curve of the
dynamical vector field is uniquely determined by the values of
these $2n-1$ functions, i.e. by the intersection of the
submanifolds $F_{i}=c_{i}$. In other words, the constants of the
motion bear all the information about the motion, up to
reparametrization. So, the search for a transformation on a vector
field that leaves invariant its integral curves up to
reparametrization is equivalent to the search for a transformation
on a vector field that does not change its constants of the
motion. Since the constants of the motion are solution of the
equation
\begin{equation}
  \mathbf{X}F_{i}=0
\end{equation}
one can look for the transformations that leave it invariant. \\
It is evident that the multiplication for a function does not
affect the solutions of the above equations: in this way one can
recover the usual reparametrization of a vector field.\\
In the quantum setting, where we mainly deal with (partial)
differential equation, the above line can serve as a guide (at
least formally). Now the equations that encode the dynamical
features of the system into accout are the eigenvalue equations%
\begin{equation}\label{eq:eigen1}
  (\hat{\mathrm{H}}-E\mathbb{I})\psi=0
\end{equation}%
As above, one can look for the transformations on these equations
that leave invariant the space of solutions (i.e. the
eigenfunctions). There can be many of them. However, for analogy
with the classical counterpart, let us restrict to the
multiplication by a function\footnote{It is worthwile to point out
explicitely that this operation is meaningful since we are dealing
with differential operators on a manifold, i.e. with a specific
representation. In the abstract setting of operators it would make
no sense to multiply by a function an abstract operator.}, that
clearly has this property. One can look for those functions $f$
such that
\begin{equation}\label{eq:eigen2}
  f(\hat{\mathrm{H}}-E\mathbb{I})\psi= (\hat{\mathrm{H}}'_{E}-\lambda_{E}\mathbb{I})\psi=0
\end{equation}
i.e. such that the eigenvalue equation for $\hat{\mathrm{H}}$
(\ref{eq:eigen1}) can be written as an eigenvalue equation for a
different operator $\hat{\mathrm{H}}'$, possibly depending on $E$.
In that case we talk of reparametrization of the differential
operator $\hat{\mathrm{H}}$, and call the operator
$\hat{\mathrm{H}}'$ that verifies eq. (\ref{eq:eigen2}) the reparametrized operator.\\
A complete analysis of this proposal and of all its implication
still has to be done and will be the subject of a subsequent
paper. Here we only point out how the motivations for it come from
the present case of study, i.e. the relationship between hydrogen
atom and harmonic oscillator: what we have done at the end of the
last section, eq. (\ref{eq:hydrogen}), (\ref{harmonic})(that is
what was usually done in previous papers in a somewhat hidden way)
is exactly what we have tried to clarify
here, putting it in a bit more general context.\\
As for the other aspect of reparametrization of a vector field,
the one connected with its completion, we notice that there is a
link between the completeness of a vector field and the
self-adjointness of the corresponding (in a sense to be specified
later) operator. This can be understood from a physical point of
view by noticing that, when dealing with Hamiltonian vector fields
that are not complete, the \A\A quantization'' of the
corresponding Hamitonian gives an Hamiltonian operator which is
not self-adjoint (some explicit examples are presented in
\cite{klauder}). In other words, what happens is that, if the
classical equation of the motion does not admit \textit{global}
solutions for arbitrary initial condition, i.e. trajectories
defined for all $t\in (-\infty, +\infty)$, the corresponding
quantum Hamiltonian will not be self-adjoint, unless one does not
impose additional (and arbitrary) conditions (and this can even
not be sufficient); this is what Klauder refers to as the \A\A
classical symptoms'' of a quantum illness (the lack of
self-adjointness).\\
This can be seen in more general and mathematically sofisticated
terms, by the means of some results by Nelson (\cite{nelson}) and
a successive application by Hunziker (\cite{huntz}; see also
\cite{mcp}, p21-22 for a concise presentation).\\
All this appears in our case of study, as mentioned in the
previuos section. There we have dealt with the Hamiltonian
$\hat{\mathrm{H}}'$ of the conformal Kepler problem which is not
self-adjoint on the Hilbert space
$\CMcal{L}^{2}(\mathbb{R}^{3},\de^{4}y)$ with $\de^{4}y$ the
Lebesgue measure on $\mathbb{R}^{4}$, and we have found a measure
$\de\mu$ with respect to with $\hat{\mathrm{H}}'$ is essentially
self-adjoint (on a suitable domain); explicitely, we had
$\de\mu=4R^{2}\de^{4}y$. This change of the measure has realized
what we are referring to as the quantum analogue of the second
aspect of reparametrization of a vector fields: namely, while in
the classical setting we multiplied the vector field by a function
in order to make it complete, in the quantum setting we
change\footnote{In our specific case, the new measure $\de\mu$ is
the old Lebesgue measure multiplied by  a function that is exactly
the same that appears in the reparametrization of the classical
vector field: it is not clear to us if this is a peculiarity of
the present problem, or a general feature} the measure in order to
make the Hamiltonian (essentially) self-adjoint.\\
In this sense the change of the measure is an echo of the
reparametrization of the dynamical vector field. However, the
whole problem needs further clarification, and we shall come back
to some of the raised issues elsewhere.
\section{Conclusions}
The main aim of this paper is to contribute to the \A\A
geometrization'' of quantum mechanics. In particular, we have
presented some ideas to deal with reduction procedure in a
systematic way. We have introduced an approach based on
differential operators, in order to preserve the geometrical
aspects of classical reduction procedure. We have considered as a
case of study the hydrogen atom, guided by the classical treatment
of the Kepler problem carried over in a previous work
(\cite{ruclass}). Using our approach supplemented by symmetry
requirements, we have arrived at a possible unfolding system and
recovered the relationship with a family of harmonic
oscillators, in analogy with the classical case.\\
Thanks to the systematic character of this procedure, we think
that this procedure could be applied to other cases of physical
interest; in particular, it would be interesting to work out this
technique to quantum mechanical systems with continuous spectrum,
in order to analyze, for example, the scattering states.\\
Moreover, there are still open issues about \A\A quantum
reparametrization'', which we hope to clarify elsewhere.
 %%%%
 %%%
 %%%%%
 %%%

\end{document}